\documentstyle[12pt]{article}

\newcommand{\ads}[1]{{\rm AdS}_{#1}}  
  
\newcommand{\be}{\begin{equation}}  
\newcommand{\ee}{\end{equation}}  
\newcommand{\beqa}{\begin{eqnarray}}  
\newcommand{\eeqa}{\end{eqnarray}}  
\newcommand{\beqar}{\begin{eqnarray*}}  
\newcommand{\eeqar}{\end{eqnarray*}}

\newcommand\gaml{\gamma_{\mu\nu}}  
\newcommand\gamu{\gamma^{\mu\nu}}  
\newcommand\Tu{T^{\mu\nu}}

\begin{document}  
  
\rightline{HUTP-99/A002, EFI-99-6}  
\rightline{NSFITP-98-132, hep-th/9902121}  
\vskip 1cm  
\centerline{\Large \bf A Stress Tensor For Anti-de Sitter Gravity}  
\vskip 1cm  
  
\renewcommand{\thefootnote}{\fnsymbol{footnote}}  
\centerline{{\bf Vijay  
Balasubramanian${}^{1,2}$\footnote{vijayb@pauli.harvard.edu} and  
Per Kraus${}^{3}$\footnote{pkraus@theory.uchicago.edu},}}   
\vskip .5cm  
\centerline{${}^1$\it Lyman Laboratory of Physics, Harvard University,}  
\centerline{\it Cambridge, MA 02138, USA}  
\vskip .5cm  
\centerline{${}^2$ \it Institute for Theoretical Physics,}  
\centerline{\it University of California,}  
\centerline{\it Santa Barbara, CA 93106, USA}  
\vskip .5cm  
\centerline{${}^3$ \it Enrico Fermi Institute,}  
\centerline{University of Chicago,}  
\centerline{\it Chicago, IL 60637, USA}  
  
\setcounter{footnote}{0}  
\renewcommand{\thefootnote}{\arabic{footnote}}  
  
\begin{abstract}  
We propose a procedure for computing the boundary stress tensor  
associated with a gravitating system in asymptotically anti-de Sitter  
space.  Our definition is free of ambiguities encountered by previous  
attempts, and correctly reproduces the masses and angular momenta of  
various spacetimes.  Via the AdS/CFT correspondence, our classical  
result is interpretable as the expectation value of the stress tensor  
in a quantum conformal field theory.  We demonstrate that the  
conformal anomalies in two and four dimensions are recovered.  The two  
dimensional stress tensor transforms with a Schwarzian derivative and  
the expected central charge.  We also find a nonzero ground state  
energy for global AdS$_5$, and show that it exactly matches the  
Casimir energy of the dual ${\cal N}=4$ super Yang-Mills theory on  
$S^3 \times R$.   
\end{abstract}

\setcounter{footnote}{0}

\section{Introduction}   
\baselineskip 0.60cm  
  
In a generally covariant theory it is unnatural to assign a local  
energy-momentum density to the gravitational field.  For instance,  
candidate expressions depending only on the metric and its first  
derivatives will always vanish at a given point in locally flat  
coordinates.  Instead, we can consider a so-called ``quasilocal  
stress tensor'', defined locally on the {\em boundary} of a given  
spacetime region.  Consider the gravitational action   
thought of as a functional of the boundary metric $\gaml$.  The  
quasilocal stress tensor associated with a spacetime region  
 has been defined by Brown and  
York to be~\cite{brownyork}: \be \Tu = {2 \over \sqrt{-\gamma}}  
{\delta S_{{\rm grav}} \over \delta \gaml}   
\label{quasi}  
\ee   
The resulting stress tensor typically diverges as the boundary is  
taken to infinity.  However, one is always free to add a boundary term  
to the action without disturbing the bulk equations of motion.  To  
obtain a finite stress tensor, Brown and York propose  a subtraction  
derived by embedding a boundary with the same intrinsic metric $\gaml$  
in some reference spacetime, such as flat space. This prescription  
suffers from an important drawback: it is not possible to embed a  
boundary with an arbitrary intrinsic metric in the reference spacetime.    
Therefore, the Brown-York procedure is generally not  well defined.

For asymptotically anti-de Sitter (AdS) spacetimes, there is an  
attractive resolution to this difficulty.  A duality has been  
proposed which  equates the gravitational action of the bulk viewed as  
a functional of boundary data, with the quantum effective action of a  
conformal field theory (CFT) defined on the AdS  
boundary~\cite{juanads,gkp,holowit}.  According to this  
correspondence, (\ref{quasi}) can be interpreted as giving the   
expectation value of the stress tensor in the  
CFT:\footnote{See~\cite{navarro,emilconf,hyunetal,horitzh} for some  
interesting  examples.}    
\be   
\langle\Tu \rangle = {2 \over \sqrt{-\gamma}} {\delta S_{{\rm  
eff}} \over \delta \gaml}.    
\label{quasi2}  
\ee   
The divergences which appear as the boundary is moved to infinity  
are then simply the standard ultraviolet divergences of quantum field  
theory, and may be removed by adding {\em local} counterterms to the  
action.  These subtractions  depend only on the intrinsic  
geometry of the boundary and are defined once and for all, in contrast  
to the ambiguous prescription involving embedding the boundary in a  
reference spacetime.  This interpretation of divergences was first   
discussed in \cite{holowit}, and has been applied to various computations  
in, e.g., \cite{henskend,hyunetal,chalmers,odint}.

Inspired by the proposed correspondence, we develop a new procedure  
for defining the stress tensor of asymptotically locally  anti-de Sitter  
spacetimes.  We renormalize the stress-energy of gravity by adding a  
finite series in boundary curvature invariants to the action.  The  
required terms are fixed essentially uniquely by requiring finiteness  
of the stress tensor.  We then show that we correctly reproduce the  
masses and angular momenta of various asymptotically AdS spacetimes  
See, {\em e.g.},~\cite{abbdes,ashmag,henn,mannetal,horhawk,hormyers} for  
previous  studies of  energy in AdS

According to (\ref{quasi2}), our definition should also exhibit the  
properties of a stress tensor in a quantum CFT.  The boundary stress  
tensor of $\ads{3}$ is expected to transform under diffeomorphisms as  
a tensor plus a Schwarzian derivative.  We verify this transformation  
rule, and so derive the existence of a Virasoro algebra with central  
charge $c = 3\ell/2G$, in agreement with the result of Brown and  
Henneaux~\cite{brownhen}.  We also demonstrate that the stress tensor  
acquires the correct trace anomaly $T^\mu_\mu = - {c \over 24\pi}{\cal  
R}$.  
  
The candidate dual to $\ads{5}$ gravity is four-dimensional ${\cal  
N}=4$ super Yang-Mills theory.  Our procedure for computing the  
spacetime stress tensor (\ref{quasi}) reproduces the expected trace  
anomaly of the gauge theory.  An interesting --- and at first  
surprising --- feature of our stress tensor is that it is generally  
non-vanishing even when the bulk geometry is exactly AdS.  In  
particular, global $\ads{5}$, with an $S^3 \times R$ boundary, has a  
positive mass.  In contrast, the reference spacetime approach, by  
construction, gives pure AdS a vanishing mass.  Our result is  
beautifully explained via the proposed duality with a boundary CFT.  
The dual super Yang-Mills theory on a sphere has a Casimir energy that  
precisely matches our computed spacetime mass.

We conclude by discussing prospects for defining an analogous quasilocal  
stress tensor in asymptotically flat spacetimes.  
  
\section{Defining The Stress Tensor}  
\label{sec:quasi}  
  
Brown and York's definition of the quasilocal stress tensor is  
motivated by Hamilton-Jacobi theory~\cite{brownyork}.   
The energy of a point  
particle is  the variation of the action with respect  
to time: $E = - \partial S / \partial t$.  In gravity, lengths are  
measured by the metric, so time is naturally replaced by the boundary  
metric $\gaml$, yielding a full stress tensor $\Tu$:   
\be   
\Tu = {2  
\over \sqrt{-\gamma}} {\delta S \over \delta \gaml}.   
\label{quasi3}  
\ee   
Here $S = S_{{\rm grav}}(\gaml)$ is the  gravitational  
action viewed as a functional of $\gaml$.  Of course, this is also the  
standard formula for the stress tensor of a field theory with action  
$S$ defined on a surface with metric $\gaml$.  
  
The gravitational action with cosmological constant   
$\Lambda = -d(d-1)/2 \ell^2$ is\footnote{Our conventions are those of
\cite{weinberg}.  Comparison with other references and certain symbolic
manipulation packages may require a flip in the sign of the Riemann
tensor.}:  
\be  
S = -{1 \over 16\pi G} \int_{\cal M} \! d^{d+1}x \, \sqrt{g} \left( R  
- {d (d-1) \over \ell^2} \right)  
- {1 \over 8\pi G} \int_{\partial {\cal M}} \! d^dx \, \sqrt{-\gamma}\,\Theta  
+ {1 \over 8 \pi G} S_{ct}(\gaml).  
\label{gaction}  
\ee  
The second term is required for a well defined variational principle  
(see, e.g.,~\cite{Wald}), and $S_{ct}$ is the counterterm action that  
we will add in order to obtain a finite stress tensor.  $\Theta$ is  
the trace of the extrinsic curvature of the boundary, and is defined  
below.

Consider foliating the $d+1$ dimensional spacetime ${\cal M}$ by a  
series of $d$ dimensional timelike surfaces homeomorphic to the  
boundary $\partial {\cal M}$.  We let $x^\mu$ be coordinates spanning  
a given timelike surface, and let $r$ be the remaining coordinate.  It  
is convenient to write the spacetime metric in an ADM-like  
decomposition~\cite{Wald}:  
\be  
ds^2 = N^2 dr^2 + \gaml(dx^\mu +N^\mu dr)(dx^\nu +N^\nu dr).  
\label{ADM}  
\ee Here $\gaml$ is a function of all the coordinates, including $r$.  
We will refer to the surface at fixed $r$ as the boundary  
$\partial{\cal M}_r$ to the interior region ${\cal M}_r$. The metric  
on $\partial{\cal M}_r$ is $\gaml$ evaluated at the boundary value of  
$r$. In AdS, the boundary metric acquires  
an infinite Weyl factor as we take $r$ to infinity. So we will more properly think of the AdS boundary as a conformal class of boundaries (see,  
e.g.,~\cite{holowit}).

To compute the quasilocal stress tensor for the region ${\cal M}_r$ we  
need to know the variation of the gravitational action with respect to  
the boundary metric $\gaml$.\footnote{See~\cite{brownyork} for a  
detailed development of the formalism.}  In general, varying the  
action produces a bulk term proportional to the equations of motion  
plus a boundary term.  Since we will always consider solutions to the  
equations of motion, only the boundary term contributes:  
\be  
\delta S =  \int_{\partial {\cal M}_r} \! d^dx \, \pi^{\mu\nu} \delta \gaml  
+ {1 \over 8 \pi G} \int_{\partial {\cal M}_r} \! d^dx   
{\delta S_{ct} \over \delta \gaml} \delta \gaml,  
\ee  
where $\pi^{\mu\nu}$ is the momentum conjugate to $\gaml$ evaluated at  
the boundary:  
\be  
\pi^{\mu\nu} = {1 \over 16 \pi G} \sqrt{-\gamma}(\Theta^{\mu\nu} -\Theta\gamu).  
\ee  
Here the extrinsic curvature is  
\be  
\Theta^{\mu\nu} = - {1 \over 2} (\nabla^\mu  
\hat{n}^\nu + \nabla^\nu \hat{n}^\mu ),  
\ee  
where $\hat{n}^\nu$ is the outward pointing normal vector to the  
boundary $\partial {\cal M}_r$. The quasilocal stress tensor is thus  
\be  
T^{\mu\nu} = {1 \over 8\pi G} \left[ \Theta^{\mu\nu} - \Theta \gamu  
+ {2 \over \sqrt{-\gamma}} {\delta S_{ct} \over \delta \gaml} \right].  
\label{renT}  
\ee $S_{ct}$ must be chosen to cancel divergences that arise as  
$\partial{\cal M}_r$ tends to the AdS boundary $\partial{\cal M}$.  In  
this limit we expect to reproduce standard computations of the mass of  
asymptotically AdS  
spacetimes~\cite{abbdes,mannetal,ashmag,horhawk,hormyers}.  Brown and  
York propose to embed $\partial {\cal M}_r$ in a pure AdS background  
and to let $S_{ct}$ be the action of the resulting spacetime region.  
A similar reference spacetime approach is taken by the authors  
of~\cite{mannetal,horhawk,hormyers}. However, as noted by all these  
authors, it is not always possible to find such an embedding, and so  
the prescription is not generally well-defined.  A reference spacetime  
is also implicitly present in the treatment of Abbott and  
Deser~\cite{abbdes} which constructs a Noether current for  
fluctuations around pure AdS.  Finally, Ashtekar and  
Magnon~\cite{ashmag} exploit the conformal structure of asymptotically  
AdS spaces to directly compute finite conserved charges.  It would be  
interesting to understand the relation of our work to their approach.

We propose an alternative procedure: take $S_{ct} $ to be a local  
functional of the intrinsic geometry of the boundary, chosen to cancel  
the $\partial{\cal M}_r \rightarrow \partial{\cal M}$ divergences in  
(\ref{renT}).  Here we set $S_{ct} = \int_{\partial {\cal M}_r} L_{ct}$,  
and state our results for AdS$_3$, AdS$_4$, and AdS$_5$:     
\beqa  
{\rm AdS}_3: \quad L_{ct} &=& - {1 \over \ell}   
\sqrt{-\gamma} \quad\quad  
\quad \quad\quad\quad\,  
\Rightarrow  \quad T^{\mu\nu} = {1 \over 8 \pi G}   
\left[\Theta^{\mu\nu} - \Theta \gamu - {1 \over \ell} \gamu \right] \nonumber \\  
{\rm AdS}_4: \quad  
L_{ct} &=& - {2 \over \ell}   
\sqrt{-\gamma} \left(1 - {\ell^2 \over 4} R\right)   
 \quad\,  
\Rightarrow \quad T^{\mu\nu} = {1 \over 8 \pi G}   
\left[\Theta^{\mu\nu} - \Theta \gamu - {2 \over \ell} \gamu   
- \ell G^{\mu\nu}  
\right] \nonumber \\  
{\rm AdS}_5: \quad  L_{ct} &=& - {3 \over \ell}   
\sqrt{-\gamma} \left(1 -  
{ \ell^2 \over 12}   R \right)  
 \quad  
\Rightarrow \quad\! T^{\mu\nu} = {1 \over 8 \pi G}   
\left[\Theta^{\mu\nu} - \Theta \gamu - {3 \over \ell} \gamu - {\ell \over 2}  
G^{\mu\nu}  
\right] \nonumber \\  
\label{results}  
\eeqa  
All tensors above refer to the boundary metric $\gaml$, and  
$G_{\mu\nu} = R_{\mu\nu} - {1 \over 2} R\gaml$ is the  
Einstein tensor of $\gaml$.     
  
As we will see, the terms appearing in $S_{ct}$ are fixed essentially  
uniquely by requiring cancellation of divergences. The number of  
counterterms required grows with the dimension of AdS space. In  
general, we are also free to add terms of higher mass dimension to  
the counterterm action for $\ads{d+1}$.  But when $d$ is odd,  
dimensional analysis shows that these terms make no contribution to  
$T^{\mu\nu}$ as the boundary is taken to infinity.  For $d$ even  
there is one potential ambiguity which we will explain and exorcise in  
later sections.  The addition of $S_{{\rm ct}}$ does not affect the  
bulk equations of motion or the Gibbons-Hawking black hole entropy  
calculations because the new terms are intrinsic invariants of the  
boundary.  
  
After adding the counterterms (\ref{results}), the stress tensor  
(\ref{renT}) has a well defined limit as $\partial{\cal M}_r  
\rightarrow \partial{\cal M}$.  (More precisely, dimensional analysis  
determines the scaling of the stress tensor with the diverging Weyl  
factor of the boundary metric.  However, observables like mass and  
angular momentum will be $r$ independent.)  To assign a mass to an  
asymptotically AdS geometry, choose a spacelike surface $\Sigma$ in  
$\partial {\cal M}$ with metric $\sigma_{ab}$, and write the boundary  
metric in ADM form:  
\begin{equation}  
\gaml dx^\mu dx^\nu = -N_{\Sigma}^{\,2}dt^2+\sigma_{ab}  
(dx^a + N_{\Sigma}^a dt)(dx^b +N_{\Sigma}^bdt).  
\label{ADM2}  
\end{equation}  
Then let $u^{\mu}$ be the timelike unit normal to $\Sigma$.  $u^\mu$  
defines the local flow of time in $\partial {\cal M}$.  If $\xi^\mu$ is a  
Killing vector generating an isometry of the boundary geometry,  
there should be an associated conserved charge.  Following Brown and  
York~\cite{brownyork}, this charge is:  
\begin{equation}  
Q_\xi = \int_\Sigma d^{d-1}x \sqrt{\sigma} \, (u^\mu \, T_{\mu\nu} \,  
\xi^\nu)  
\label{conserved}  
\end{equation}  
The conserved charge associated with time translation is then the mass  
of spacetime.   Alternatively, we can define a proper energy density   
\be  
\epsilon = u^\mu u^\nu T_{\mu\nu}.  
\ee  
To convert to mass, multiply by the lapse $N_\Sigma$ appearing in   
(\ref{ADM2}) and integrate:  
\be  
M = \int_{\Sigma} \! d^{d-1}x \, \sqrt{\sigma}\, N_\Sigma\, \epsilon.  
\label{themass}  
\ee  
This definition of mass coincides with the conserved quantity in  
(\ref{conserved}) when the timelike Killing vector is $\xi^\mu =  
N_\Sigma \, u^\mu$. Similarly, we can define a momentum  
\be  
P_{a} =\int_{\Sigma} \! d^{d-1}x \, \sqrt{\sigma}\, j_{a},  
\label{themom}  
\ee  
where  
\be  
j_a = \sigma_{ab} u_{\mu} T^{a\mu}.  
\ee  
When $a$ is an angular direction, $P_a$ is the corresponding angular   
momentum.     
  
Although we have only written the gravitational action in  
(\ref{gaction}), our formulae are equally valid in the presence of  
matter.  In particular, (\ref{themass}) and (\ref{themom}) give the total  
mass and momentum of the entire matter plus gravity system.

\section{AdS$_3$}  
\label{ads3}  
  
We begin with the relatively simple case of AdS$_3$. We will show that  
our prescription correctly computes the mass and angular momentum of  
BTZ black holes, and reproduces the transformation law and conformal  
anomaly of the stress tensor in the dual CFT.  
  
The Poincar\'e  
patch of AdS$_3$ can be written as:\footnote{See, e.g.,~\cite{bthz}  
for the embedding of the Poincar\'e patch in global  $\ads{3}$.}   
\be  
ds^2 = {\ell^2 \over r^2} dr^2 + {r^2 \over \ell^2} (-dt^2 +dx^2).  
\label{poinads3}  
\ee  
A boundary at fixed $r$ is  conformal to $R^{1,1}$: $-\gamma_{tt}=  
\gamma_{xx} = r^2/\ell^2$.    The normal vector to surfaces of constant  
$r$ is  
\be  
\hat{n}^\mu = {r \over \ell} \delta^{\mu,r}.  
\ee  
Applying (\ref{renT}) we find   
\beqa  
8 \pi G \,T_{tt} &=& -{r^2 \over \ell^3} + {2 \over \sqrt{-\gamma}}  
{\delta S_{ct} \over \delta \gamma^{tt}}  \nonumber  \\  
8 \pi G \,T_{xx} &=& {r^2 \over \ell} + {2 \over \sqrt{-\gamma}}  
{\delta S_{ct} \over \delta \gamma^{xx}}  \nonumber  \\  
8 \pi G \,T_{tx} &=&  {2 \over \sqrt{-\gamma}}  
{\delta S_{ct} \over \delta \gamma^{tx}}.      
\eeqa  
Neglecting $S_{ct}$, one would obtain divergent results for physical   
observables such as the mass  
\be  
M = \int \! dx\, \sqrt{g_{xx}} \, N_\Sigma \, u^t u^t \, T_{tt} =   
\int \! dx\, T_{tt}  ~ \sim ~ r^2 ~ \rightarrow ~ \infty.  
\label{divm}  
\ee  
So $T_{tt}$ must be independent of $r$ for large $r$  in order for the  
spacetime to have a finite mass density.     
  
$S_{ct}$ is defined essentially uniquely by the requirement that it be  
a local, covariant function of the intrinsic geometry of the boundary.  
It is readily shown that the only such term that can cancel the  
divergence in (\ref{divm}) is $S_{ct} = (-1/\ell) \int  
\sqrt{-\gamma}$. This then yields $T_{\mu\nu} =0$, which is clearly  
free of divergences. In general, we could have added further higher  
dimensional counterterms such as $R$ and $R^2$.  Dimensional  
analysis shows that terms higher than $R$ vanish too rapidly at  
infinity to contribute to the stress tensor.  The potential  
contribution from the metric variation of $R$ is $G^{\mu\nu}$, the  
Einstein tensor, which vanishes identically in two dimensions.  So the  
minimal counterterm in (\ref{results}) completely defines the  
$\ads{3}$ stress tensor.

Since the stress tensor is now fully specified, it must reproduce the  
mass and angular momentum of a known solution. To check this, we study  
spacetimes of the form:  
\be  
ds^2 =  {\ell^2 \over r^2} dr^2 + {r^2 \over \ell^2} (-dt^2 +dx^2)  
+ \delta g_{MN}dx^M dx^N.  
\ee  
Working to first order in $\delta g_{MN}$, we find   
\begin{eqnarray}  
8 \pi G \,T_{tt} &=&  
 {r^4 \over 2 \ell^5} \delta g_{rr} + {\delta g_{xx} \over \ell}  
-{r \over 2 \ell} \partial_r \delta g_{xx} \nonumber \\  
8 \pi G \,T_{xx} &=& {\delta g_{tt}\over \ell}  
-{ r \over 2 \ell } \partial_r \delta g_{tt} -  
{r^4 \over 2 \ell^5} \delta g_{rr}  \nonumber \\  
8 \pi G \,T_{t x} &=& {1 \over \ell} \delta g_{tx}  
-{ r \over 2 \ell} \partial_r \delta g_{tx}   
\end{eqnarray}  
The mass and momentum are:  
\beqa  
M &=& {1 \over 8 \pi G} \int \! dx \,  
\left[{r^4 \over 2 \ell^5} \delta g_{rr} + {\delta g_{xx} \over \ell}  
-{r \over 2 \ell} \partial_r \delta g_{xx} \right] \nonumber \\  
P_x &=& -{1 \over 8 \pi G} \int \! dx \,  
\left[{1 \over \ell} \delta g_{tx}  
-{ r \over 2 \ell} \partial_r \delta g_{tx}\right].  
\label{MP}   
\eeqa  
  
We can apply these formulae to the spinning BTZ solution~\cite{BTZ,bthz}:  
\be  
ds^2 = -N^2 dt^2 + \rho^2(d\phi+ N^\phi dt)^2+{r^2 \over N^2 \rho^2} dr^2  
\ee  
with  
\beqa  
N^2 &=& {r^2(r^2-r_+^2) \over \ell^2 \rho^2}, ~~~~~~~~~~~~  
N^\phi = - {4GJ \over \rho^2}, \nonumber \\  
\rho^2 &=& r^2 +4GM\ell^2 -{1 \over 2} r_+^2, ~~~~~~~~~~~~  
r_+^2 = 8G\ell\sqrt{M^2\ell^2-J^2},  
\eeqa  
where $\phi$ has period $2 \pi$.  Expanding the metric for large $r$ we find  
\be  
\delta g_{rr} = {8GM \ell^4 \over r^4},  \quad\quad  
\delta g_{tt} = 8GM, \quad\quad  \delta g_{t\phi} = -4GJ.  
\ee  
Inserting these into (\ref{MP}) with $x \rightarrow \ell\phi$ and $\int  
\!dx \rightarrow \ell \int_0^{2\pi} \! d\phi$ gives the correct relations  
$M=M$ and $P_\phi = J$ in agreement with conventional techniques.  
When $M=-1/8G$ and $J=0$, the BTZ metric reproduces global $\ads{3}$,  
while the $M=0$, $J=0$ black hole looks like Poincar\'e $\ads{3}$  
(\ref{poinads3}) with an identification of the boundary.  It may seem  
surprising that global $\ads{3}$ apparently differs in mass from the  
Poincar\'e patch.  The difference arises because the time directions  
of these coordinates do not agree, giving rise to different  
definitions of energy.  
  
\subsection{Conformal Symmetry of AdS$_3$}   
  
Brown and Henneaux~\cite{brownhen} have shown that gravity in  
asymptotically AdS$_3$ spacetime is a conformal field theory with  
central charge $c = 3\ell/2G$.  Both as a check of our approach, and  
because our covariant method will offer an alternative to the  
Hamiltonian formalism adopted in~\cite{brownhen} and the Chern-Simons  
methods of~\cite{banados}, we would like to reproduce this  
result.\footnote{Related work has been done by Hyun et.al.~\cite{hyunetal}}  
  
In light of the AdS/CFT correspondence, we can think of the conformal   
symmetry group as arising from a $1+1$ dimensional non-gravitational   
quantum field theory living (loosely speaking) on the boundary of AdS$_3$.  
On a plane with metric $ds^2 =-dx^+ dx^-$, diffeomorphisms of the form  
\be  
 x^+  \rightarrow x^+ - \xi^+(x^+), ~~~~~~~~~~~~  
 x^-  \rightarrow x^- - \xi^-(x^-)   
\label{diffeo}  
\ee  
transform the stress tensor as:  
\beqa  
T_{++} &\rightarrow&  T_{++} +(2\partial_+ \xi^+ \, T_{++}+\xi^+ \, \partial_+   
T_{++}) -{c \over 24 \pi} \, \partial_+^{\,3}\xi^+,   
     \nonumber  \\  
T_{--} &\rightarrow&  T_{--} +(2\partial_- \xi^- \, T_{--}+\xi^- \, \partial_-   
T_{--}) - {c \over 24 \pi} \, \partial_-^{\,3}\xi^-.    
\label{transform}  
\eeqa  
The terms in parenthesis are just the classical tensor transformation  
rules, while the last term is a quantum effect.  Let us briefly recall  
the origin of the latter.  Although (\ref{diffeo}) is classically a  
symmetry of the CFT, it is quantum mechanically anomalous since we  
must specify a renormalization scale $\mu$.  To obtain a symmetry  
under (\ref{diffeo}), $\mu$ must also be rescaled to have  the same  
measured value in the new coordinates as in the original  
coordinates.  Equivalently, the metric should be Weyl rescaled to  
preserve the form $ds^2=-dx^+ dx^-$.  Such a rescaling of lengths acts  
non-trivially in the quantum theory and produces the extra terms in  
(\ref{transform}).  
  
We will focus on obtaining the final terms in (\ref{transform}) by  
starting from AdS$_3$ in the form  
\be  
ds^2= {\ell^2 \over r^2} dr^2-r^2 dx^+ dx^-,  
\ee  
for which $T_{\mu\nu}=0$.   We think of the dual CFT as living on the   
surface $ds^2 =-r^2 dx^+ dx^-$ with $r$ eventually taken to infinity.   
Now consider the diffeomorphism (\ref{diffeo}).  As above, this is not  
a symmetry since it introduces a Weyl factor into the boundary metric.  
To obtain a symmetry one must leave the asymptotic form of the metric   
invariant, and the precise conditions for doing so  have been given by  
Brown and Henneaux~\cite{brownhen}:  
\beqa  
g_{+-}&=& -{r^2 \over 2} +{\cal O}(1),  ~~~~~~~  
g_{++} = {\cal O}(1), ~~~~~~~   
g_{--}= {\cal O}(1), \nonumber \\  
g_{rr}&=& {\ell^2 \over r^2} +{\cal O}({1 \over r^4}), ~~~~~~~  
g_{+r} = {\cal O}({1 \over r^3}), ~~~~~~~  
g_{-r} = {\cal O}({1 \over r^3}).    
\eeqa  
The diffeomorphisms which respect these conditions are:  
\beqa  
x^+ &\rightarrow& x^+ -\xi^+ -{\ell^2 \over 2 r^2} \partial_-^{\,2}\xi^-  
\nonumber \\  
x^- &\rightarrow& x^- -\xi^- -{\ell^2 \over 2 r^2} \partial_+^{\,2}\xi^+  
 \nonumber\\   
r &\rightarrow& r + {r \over 2} (\partial_+ \xi^+ +\partial_- \xi^-).  
\label{newdiffeo}  
\eeqa  
For large $r$, the corrections to the $x^\pm$ transformations  are subleading, and we recover   
(\ref{diffeo}).  
The metric then transforms as  
\be  
ds^2 \rightarrow {\ell^2 \over r^2} dr^2 -r^2 dx^+ dx^-   
- {\ell^2 \over 2} (\partial_+^{\,3} \xi^+)(dx^+)^2   
- {\ell^2 \over 2} (\partial_-^{\,3} \xi^-)(dx^-)^2.  
\label{newmet}  
\ee  
Since the asymptotic metric retains its form, this   
 transformation is a symmetry.   Using  (\ref{newmet}) we  
compute the stress tensor to be  
\be  
T_{++}= -{\ell \over 16\pi G}\partial_+^{\,3} \xi^+, ~~~~~~~~~~~~~~  
 T_{--}= -{\ell \over 16\pi G}\partial_-^{\,3} \xi^-.  
\ee  
This agrees with (\ref{transform}) if   
\be  
c = {3 \ell \over 2G}.  
\ee  
Thus we have verified the result of Brown and Henneaux~\cite{brownhen}.    
  
In the CFT the full transformation law arose from doing a renormalization  
group rescaling of $\mu$, while on the gravity side it arose from a  
diffeomorphism which rescaled the radial position of the boundary.  This fits  
very nicely with the general feature of the AdS/CFT correspondence that scale  
size in the CFT is dual to the radial position in AdS.  According to   
\cite{SussWitt}, $r$ specifies an effective UV cutoff in the  
CFT; by rescaling $r$ before taking it to infinity we are changing the   
way in which the cutoff is removed --- but this is just the definition of  
a renormalization group transformation.  
  
We restricted attention to the diffeomorphism  
(\ref{newdiffeo}) because we were interested in symmetries which preserved  
the form of the boundary metric.  More general  
diffeomorphisms may be studied, but these will modify the form of  
the CFT and so are not symmetries.  
  
\subsection{Conformal Anomaly for AdS$_3$}  
\label{ads3anom}  
  
The stress tensor of a $1+1$ dimensional CFT has a trace anomaly   
\be  
T^\mu_\mu = -{c \over 24\pi} R  
\label{trace}  
\ee  
We will now verify that our quasilocal stress tensor has a trace of precisely  
this form.  The mechanism for obtaining a conformal anomaly from the AdS/CFT  
correspondence was outlined by Witten \cite{holowit} and studied in detail  
by Henningson and Skenderis \cite{henskend}.  Our approach is somewhat   
different from that of \cite{henskend}.    
  
Taking the trace of the AdS$_3$ stress tensor appearing in  
(\ref{results}) we find  
\be  
T^\mu_\mu = -{1 \over 8\pi G}(\Theta + 2/\ell).  
\label{quasitrace}  
\ee  
(\ref{quasitrace}) gives the trace in terms of the extrinsic curvature;  
to compare with (\ref{trace}) we need to express the result in terms of  
the intrinsic curvature of the boundary.    
  
Since (\ref{quasitrace}) is manifestly covariant, we may compute the right  
hand side in any convenient coordinate system.  We write  
\be  
ds^2 = {\ell^2 \over r^2} dr^2 +\gaml dx^\mu dx^\nu  
\label{hsmetric}  
\ee  
  The  
extrinsic curvature in these coordinates is  
\be  
\Theta_{\mu\nu} = -{r \over 2 \ell} \partial_r \gaml.  
\label{excurv}  
\ee  
So in this coordinate system (\ref{quasitrace}) becomes  
\be  
T^\mu_\mu = - {1 \over 8\pi G}   
\left[- {r \over 2 \ell} \gamu \partial_r  
\gaml +{2 \over \ell} \right]  
\ee  
To complete the calculation we need $\gaml$ as a power series  
in $1/r$.   Einstein's equations show \cite{feff} that only  
even powers appear and that the leading term goes as $r^2$ .  So we write  
\be  
\gaml = r^2 \gaml^{(0)} + \gaml^{(2)} + \cdots.  
\ee  
There are additional higher powers of $1/r$ as well as logarithmic terms  
\cite{feff},  
but these will not be needed.  We now have  
\be  
T^\mu_\mu = - {1 \over 8\pi G} {1 \over \ell r^2}{\rm Tr}\left[  
(\gamma^{(0)})^{-1} \gamma^{(2)}\right] + \cdots.  
\label{trpert}  
\ee  
Solving Einstein's equations perturbatively gives \cite{henskend}  
\be  
{\rm Tr}\left[  
(\gamma^{(0)})^{-1} \gamma^{(2)}\right] = {\ell^2 r^2 \over 2} R  
\ee  
where $R$ is the curvature of the metric $\gaml$.  Finally, inserting  
this into (\ref{trpert}) and taking $r$ to infinity we obtain  
\be  
T^\mu_\mu = - { \ell \over 16 \pi G}{\cal R},  
\ee  
which agrees with (\ref{trace}) when $c=3\ell/2G$.  
  
\section{AdS$_{4}$}  
  
The only difference between the $\ads{4}$ and $\ads{3}$ stress tensor  
derivations is the need for an extra term in $S_{ct}$ to cancel  
divergences.  Again, start with AdS$_4$ in Poincar\'e form:  
\be  
ds^2= {\ell^2 \over r^2}dr^2 +{r^2 \over \ell^2}(-dt^2 +dx_idx_i)   
\quad\quad\quad  i=1,2.  
\label{poincare4}  
\ee  
Following Sec.~\ref{ads3}, we compute the mass of the spacetime and  
demand that it be finite:   
\be  
M = \int \! d^2x\, \sqrt{g_{xx}} N_\Sigma u^t u^t T_{tt} =    
\int \! d^2x\, {r \over \ell}T_{tt}.  
\label{mass41}  
\ee  
A finite mass density requires $T_{tt} \sim r^{-1}$ for large $r$.  
Evaluating the stress tensor for the metric (\ref{poincare4}), we find  
\beqa  
8 \pi G \,T_{tt} &=& -2{r^2 \over \ell^3} + {2 \over \sqrt{-\gamma}}  
{\delta S_{ct} \over \delta \gamma^{tt}}  \nonumber  \\  
8 \pi G \,T_{x_ix_j} &=& 2{r^2 \over \ell}\delta_{ij}   
+ {2 \over \sqrt{-\gamma}}  
{\delta S_{ct} \over \delta \gamma^{x_ix_j}}  \nonumber  \\  
8 \pi G \,T_{tx_i} &=&  {2 \over \sqrt{-\gamma}}  
{\delta S_{ct} \over \delta \gamma^{tx_i}}.      
\eeqa   
The divergences are cancelled by choosing   
$S_{ct}= -{2 \over \ell} \int \sqrt{-\gamma}$; in particular we find that  
$T_{\mu\nu}=0$.  
  
Now consider AdS$_4$ in global coordinates:  
\be  
ds^2= - \left(1+ {r^2 \over \ell^2} \right)dt^2 +   
{dr ^2 \over \left(1+ {r^2 \over \ell^2} \right)}   
+r^2 (d\theta^2+\sin^2{\theta} \, d\phi^2).  
\label{global4}  
\ee  
It is easy to show that the mass is still given by (\ref{mass41}) in  
the limit $r \rightarrow \infty$, after replacing $d^2x$ by $\sin\theta \,  
d\theta \, d\phi$.  We find that the counterterm introduced above  
correctly removes the $r^2$ divergence in $T_{\mu\nu}$, but there remains  
a $r^0$ behaviour (leading to a divergent mass  
 which can be cancelled by adding $\int \ell  
\sqrt{-\gamma}R/2$ to $S_{ct}$.  Altogether, this gives the counterterm  
action written in (\ref{results}).  We are free to add higher  
dimensional objects like $R^2$ to $S_{ct}$, but they vanish too quickly  
at the $\ads{4}$ boundary to contribute to the stress tensor.  In  
total, the stress tensor for  the metric (\ref{global4}) is:  
  
\begin{eqnarray}  
8 \pi G T_{tt} &=&{ \ell \over 4 r^2}+ \cdots  \nonumber \\  
8 \pi G T_{\theta \theta} &=& {\ell^3 \over 4 r^2}+ \cdots \\  
8 \pi G T_{\phi \phi} &=&{\ell^3 \over 4 r^2}\sin^2{\theta}+ \cdots  
\nonumber  
\end{eqnarray}  
  
We test our definition on the  AdS$_4$-Schwarzschild solution:  
\be  
ds^2=-\left[{r^2 \over \ell^2} +1-{r_0 \over r} \right]dt^2  
+ \left[{r^2 \over \ell^2} +1-{r_0 \over r}\right]^{-1}dr^2  
+r^2d\Omega_2^2.  
\label{adssch4}  
\ee  
  We find  
\be  
8 \pi G T_{tt} = {r_0 \over \ell r} + \cdots,  
\ee  
leading to a mass   
\be  
M={r_0 \over 2G}  
\label{globalmass4}  
\ee  
This agrees with the standard definition of the $\ads{4}$ black hole  
mass.   
  
\subsection{Conformal Anomaly for AdS$_4$}  
  
Direct computation shows that the stress tensor for AdS$_4$ is  
traceless.  There is also a general argument that the trace vanishes  
for any even dimensional AdS, which we give instead.   
  
The stress tensor for AdS$_{d+1}$ has length dimension $-d$.  Since  
for large $r$ the Weyl factor multiplying the boundary metric is  
proportional to $r^2$, it must be the case that  
\be  
T^\mu_\mu \sim {1 \over r^d}.  
\label{trform}  
\ee  
Working in coordinates like (\ref{hsmetric}), the trace has the structure  
\be  
 T^\mu_\mu \sim  r \gamu \partial_r \gaml + ({\rm curvature ~invariants~  
of~ } \gaml).  
\label{gentrace}  
\ee  
Now, $\gaml$ has an expansion in {\em even} powers of $r$~\cite{feff}:   
\be  
\gaml = r^2 \sum_{n =0}^{\infty} {\gaml^{(2n)} \over r^{2n}}.  
\ee  
Using this in (\ref{gentrace}), and the fact that scalar  
curvature invariants always involve even powers of the metric, we find  
that only even powers of $r$ can appear in the trace.  Comparing with  
(\ref{trform}), shows that the stress tensor must vanish for odd $d$.

This result is expected from the AdS/CFT correspondence, since even  
dimensional AdS bulk theories are dual to odd dimensional CFTs, which  
have a vanishing trace anomaly.

\section{{\bf AdS$_5$}}  
  
The $\ads{5}$ counterterms are derived in parallel with AdS$_4$, so we  
can be brief.  The expression for the spacetime mass is now:  
\be  
M = \int \! d^3x\, \sqrt{g_{xx}} \, N_\Sigma \, u^t u^t \, T_{tt} =   
\int \! d^3x\, {r^2 \over \ell^2}T_{tt}.  
\label{mass51}  
\ee  
A finite mass density therefore requires $T_{tt} \sim r^{-2}$ for large $r$.  
Upon evaluating the stress tensor in Poincar\'e and global coordinates and  
imposing finiteness, we arrive at the counterterms written in  
(\ref{results}).    By dimensional analysis, the only possible higher  
dimensional terms in $S_{ct}$ that could make a finite contribution to  
the stress tensor are the squares of the Riemann tensor, the Ricci  
tensor and the Ricci scalar of the boundary metric.  We will discuss  
these potential ambiguities in Sec.~5.1.

We now check our definition against the known mass of particular solutions.  
Consider the metric  
\be  
ds^2={r^2 \over \ell^2}\left[-\left(1-{r_0^4 \over r^4}\right)dt^2  
+ (dx_i)^2\right]+\left(1-{r_0^4 \over r^4}\right)^{-1}{\ell^2 \over r^2}dr^2  
\ee  
that arises in the near-horizon limit of the D3-brane (see,  
e.g.,~\cite{hormyers}).   The stress tensor is    
\beqa  
8\pi G T_{tt}&=&{3 r_0^4 \over 2 \ell^3 r^2} + \cdots  \nonumber \\  
8\pi G T_{x_i x_i}&=&{ r_0^4 \over 2 \ell^3 r^2} + \cdots.  
\eeqa  
Using (\ref{mass51}) gives  
\be  
M = {3 r_0^4 \over 16 \pi G \ell ^5} \int \! d^3x.  
\ee  
This agrees with the standard formula for the mass density of this solution~\cite{hormyers}.

Next, consider the AdS-Schwarzschild black hole solution,  
\be  
ds^2=-\left[{r^2 \over \ell^2} +1-\left({r_0 \over r}\right)^2 \right]dt^2  
+ {dr^2 \over \left[{r^2 \over \ell^2} +1-\left({r_0 \over r}\right)^2 \right]}  
+r^2(d\theta^2 + \sin^2{\theta}d\phi^2 + \cos^2{\theta} d\psi^2).  
\ee  
Note that  $r_0=0$ gives the global AdS$_5$ metric.  We find  
\beqa  
8 \pi G T_{tt} &=& {3 \ell \over 8 r^2} +{3 r_0^2 \over 2 \ell r^2} + \cdots,  
\nonumber \\  
8 \pi G T_{\theta\theta} &=& { \ell^3 \over 8 r^2}   
+{\ell r_0^{\,2} \over 2  r^2} + \cdots,  
\nonumber \\  
8 \pi G T_{\phi\phi} &=& \left({ \ell^3 \over 8 r^2}   
+{\ell r_0^{\,2} \over 2  r^2}\right)\sin^2{\theta} + \cdots, \nonumber \\  
8 \pi G T_{\psi\psi} &=& \left({ \ell^3 \over 8 r^2}   
+{\ell r_0^{\,2} \over 2  r^2}\right)\cos^2{\theta} + \cdots,  
\eeqa  
The mass is  
\be  
M={3 \pi \ell^2 \over 32  G} +  
{3 \pi r_0^2 \over 8  G}.  
\label{globalmass5}  
\ee  
The standard mass of this solution is $3 \pi r_0^2/8  
G$~\cite{hormyers}, which is the second term of our result  
(\ref{globalmass5}).  We have the additional constant  
$3\pi\ell^2/32G$ which is then the  mass of pure global $\ads{5}$  
when $r_0=0$.  It seems unusual from the gravitational point of view  
to have a mass for a solution that is a natural vacuum, but we will  
show that this is precisely correct from the perspective of the  
AdS/CFT correspondence.

\vskip 0.6cm  
\noindent  
{\em \bf Casimir Energy}\footnote{We thank Gary Horowitz for pointing  
out the relevance of the CFT Casimir energy to our result, and fo  
discussing his related work with Hirosi Ooguri.}  
\vskip 0.2cm  
  
String theory on $\ads{5} \times S^5$ is expected to be dual  to four  
dimensional ${\cal N}=4$, $SU(N)$ super Yang-Mills~\cite{juanads}.  
We use the conversion formula to gauge theory  
variables:   
\be  
{\ell^3 \over G} ={2 N^2 \over \pi}.  
\label{conversion}  
\ee  
Then, setting $r_0=0$, the mass of global AdS$_5$  is:  
\be  
M = {3 N^2 \over 16 \ell}.  
\label{gmass5}  
\ee  
The  Yang-Mills dual of $\ads{5}$ is defined on the global $\ads{5}$  
boundary with topology $S^3 \times R$.  A quantum field theory on suc  
a manifold   can have a nonvanishing vacuum energy --- the  
Casimir effect.  In the   
free field limit, the Casimir energy on $S^3 \times R$  
is:\footnote{Noting that $S^3 \times R$ is the Einstein static  
universe, we can adopt the results of~\cite{birrdav}.}   
\be  
E_{ {\rm casimir}} = {1 \over 960 r}(4n_0 +17 n_{1/2} +88n_1),  
\label{casimir}  
\ee  
where $n_0$ is the number of real scalars, $n_{1/2}$ is the number of Weyl  
fermions, $n_1$ is the number of gauge bosons, and $r$ is the radius of   
$S^3$.  For SU(N), ${\cal N}=4$ super Yang-Mills $n_0=6(N^2-1)$, $n_{1/2}=4(N^2-1)$ and $n_1 =N^2-1$ giving:  
\be  
E_{ {\rm casimir}} = {3(N^2 -1) \over 16 r}.  
\ee  
To compare with (\ref{gmass5}), remember that $M$ is measured with  respect to coordinate time while the Casimir energy is defined with respect to proper boundary time. Converting to coordinate time by multiplying by $\sqrt{-g_{tt}} = r/\ell$ gives the Casimir ``mass":  
\begin{equation}  
M_{ {\rm casimir}} = {3(N^2 -1) \over 16 \ell}.  
\end{equation}  
In the large $N$ limit we find precise agreement with the gravitational mass  
(\ref{gmass5}) of global $\ads{5}$.     
  
 In related work, Horowitz and Myers \cite{hormyers}   
compared the mass of an analytically continued  
non-extremal D3-brane solution to the corresponding free-field Casimir energy  
in the gauge theory, and found agreement up to an overall factor of  
$3/4$.  They argued that the mathematical origin of the discrepancy  
was the same as for a $3/4$ factor~\cite{gubser} relating the gravitational entropy  
of the system to a free field entropy computation in the CFT dual.  In  
both cases, the gravitational result is valid at strong gauge coupling  
and, apparently, the extrapolation from the free limit of the gauge  
theory involves a factor of $3/4$.

In our case, however, the coefficients match precisely.  In general,  
 gravity calculations may not be extrapolated to the weakly  
coupled gauge theory, because large string theoretic corrections can  
deform the bulk geometry in this regime.  This is the origin of the  
$3/4$ factor discussed above.  In our case, pure $\ads{5}$ is  
protected from stringy corrections because all tensors which might  
modify Einstein's equation actually vanish when evaluated in this  
background~\cite{kallraj}.  This is why the Casimir energy in the  
weakly coupled, large $N$ Yang-Mills exactly   
matches the gravitational mass of  
spacetime.   
  
\subsection{Conformal Anomaly for AdS$_5$}  
  
The $\ads{5}$ conformal anomaly computation is a more   
laborious version of the $\ads{3}$ result in  Sec.~\ref{ads3anom}.  
The trace of the $\ads{5}$ stress tensor in (\ref{results}) is  
\be  
T^\mu_\mu = -{1 \over 8\pi G}(3\Theta+12/\ell -\ell R/2).  
\ee  
Again, write the bulk metric in the form (\ref{hsmetric}) so that   
(\ref{excurv}) gives the extrinsic curvature, yielding  
\be  
T^\mu_\mu = - {1 \over 8\pi G}\left[-{3r \over 2\ell} \gamu \partial_r \gaml  
+{12 \over \ell} - {\ell \over 2}  R(\gaml)\right].  
\label{anom5a}  
\ee  
To identify the anomaly we must compute $\gaml$ to order $r^{-2}$:  
\be  
\gaml = r^2 \gaml^{(0)} + \gaml^{(2)} + r^{-2}\gaml^{(4)}+ \cdots.  
\ee  
The coefficients are found  to be~\cite{henskend}  
\beqa  
\gaml^{(2)} &=& {\ell^2 \over 2} (R^{(0)}_{\mu\nu} - {1 \over 6}  
R^{(0)}\gaml^{(0)} ) \nonumber  \\  
{\rm Tr}  
\left[(\gamma^{(0)})^{-1} \gamma^{(4)}\right] &=&   
{1 \over 4} {\rm Tr}\left[\left((\gamma^{(0)})^{-1} \gamma^{(2)}\right)^2\right].  
\eeqa  
We also need the expansion of ${\cal R}(\gaml)$:  
\be  
R(\gaml) = {1 \over r^2} R^{(0)}   
+  {\delta R \over \delta \gaml}|_{r^2\gaml^{(0)}}  
\gaml^{(2)}   
= {1 \over r^2}R^{(0)}   
- { \ell^2 \over 2 r^4} \left(R_{(0)}^{\mu\nu}R^{(0)}_{\mu\nu}  
-{1 \over 6} R_{(0)}^2 \right).  
\ee  
Inserting these results into (\ref{anom5a}) and doing some algebra, one finds  
\be  
T^\mu_\mu = - {\ell^3 \over 8\pi G}\left[    
-{1 \over 8}R^{\mu\nu}R_{\mu\nu} + {1 \over 24}R^2 \right].  
\label{anom5}  
\ee  
This result for the trace agrees with the work of  Henningson and Skenderis~\cite{henskend}.  These authors also show that upon using (\ref{conversion}), precise agreement is obtained with the  
conformal anomaly of ${\cal N}=4$ super Yang-Mills.  
  
\vskip 0.6cm  
\noindent  
{\em \bf An Ambiguity}  
\vskip 0.2cm  
  
The minimal $\ads{5}$ counterterm action in (\ref{results}) can be  
augmented by the addition of terms quadratic in the Riemann tensor,  
Ricci tensor and Ricci scalar of the boundary metric.\footnote{Higher  
dimensional invariants give a vanishing contribution to the stress  
tensor at the AdS boundary.}   A convenient basis for this ambiguity is  
provided by:   
\begin{equation}  
\Delta S_{ct} = \ell^3\int_{\partial {\cal M}_r}\! d^4x \sqrt{-\gamma} \left[a\, E + b  
\,C_{\mu\nu\rho\sigma}  C^{\mu\nu\rho\sigma} + c \, R^2 \right].  
\label{ambact}  
\end{equation}  
The first term is the Euler invariant $E = R_{\mu\nu\rho\sigma}  
R^{\mu\nu\rho\sigma} - 4 R_{\mu\nu} R^{\mu\nu} + R^2$   
and vanishes under variation, so we can omit it without loss of  
generality. $C^{\mu\nu\rho\sigma}$ is the Weyl tensor.  Varying  
$\Delta S_{ct}$ with respect to the boundary metric produces an ambiguity  
in the  
stress tensor:  
\begin{equation}  
\Delta T_{\mu\nu} = \left( {\ell^3 \over 16\pi G} \right)  
(b \, H_{\mu\nu}^b + c\, H_{\mu\nu}^c).  
\,   
\label{ambstress}  
\end{equation}  
The tensors $H^b$ and $H^c$ are computed in~\cite{birrdav}; their trace  
gives a contribution to the anomaly  
\begin{equation}  
\Delta T^\mu_\mu \propto  \Box R.  
\label{tramb}  
\end{equation}  
For general boundary metrics there is therefore a two parameter set of  
possible stress tensors, whose anomalies have varying coefficients for  
$\Box R$.  Exactly the same ambiguity is  present in the  
definition of the  renormalized stress tensor of the dual field theory  
on the curved boundary~\cite{birrdav}.  Our gravitational result  
can only be matched to field theory computations after the ambiguous  
parameters are matched.  
For conformally flat boundaries the tensor $H^b_{\mu\nu}$ vanishes  
leaving a one parameter ambiguity, which is fully specified by the  
coefficient of $\Box R$ in the anomaly.  So we learn from  
(\ref{anom5}) that gravitational energies computed with the minimal  
counterterm action $\Delta S_{ct}=0$ should be compared with a field  
theory regularization which produces a vanishing $\Box R$ anomaly  
coefficient.  Precisely this was done in the above comparison of  
Casimir energies for global AdS$_5$.  The boundary $S^3 \times R$ is  
conformally flat, and we have checked that the field theory  
computation that produces (\ref{casimir}) yields no $\Box R$ term in  
the anomaly.  This explains the agreement between the gravity and  
field theory results, despite the apparent ambiguity in choosing  
$\Delta S_{ct}$  
\section{Discussion}  
  
We have formulated a stress tensor which gives a well-defined meaning  
to the notions of energy and momentum in AdS.  
Through the AdS/CFT  
correspondence, we have also found results for the expectation value  
of the stress tensor in the dual CFT.  Our proposal exhibits the  
desired features of a stress tensor, both from the gravitational and  
CFT points of view.  
  
The procedure we have followed for defining the stress tensor is a  
particular example of the ideas developed in~\cite{probes}.  There it  
was shown how to associate the asymptotic behavior of each bulk field  
with the expectation value of a CFT operator.  The relation studied  
here between the gravitational field and the stress tensor is an  
example of this correspondence.  
  
It would be desirable to formulate an analogous stress tensor in  
asymptotically flat spacetimes.  It is not immediately clear how to  
define counterterms, since there is no longer a dimensionful parameter  
like $\ell$ allowing one to form a dimensionless counterterm action.  
On the other hand, flat spacetime is recovered from AdS by taking  
$\ell \rightarrow \infty$, so we might expect that applying this  
limit to our formulae would yield the appropriate stress tensor.  
However, the situation is complicated by the fact that we must keep  
$r$ finite while applying the limit, taking $r \rightarrow \infty$  
afterwards.  The stress tensor at finite $r$ should be interpreted in  
a CFT with an ultraviolet cutoff~\cite{SussWitt}. This implies that  
the limits $\ell \rightarrow  \infty$, $r \rightarrow \infty$ can be  
understood in renormalization  group terms~\cite{toappear}.

\vspace{0.2in}  
  
\paragraph{Acknowledgments: }   
V.B. is supported by the Harvard Society of Fellows and NSF grants  
NSF-PHY-9802709 and NSF-PHY-9407194.  P.K. is supported by NSF Grant  
No. PHY-9600697. We have had helpful discussions with Emil Martinec,  
Joe Polchinski, Jennie Traschen and, particularly, Gary Horowitz and  
Don Marolf.



\begin{thebibliography}{10}  
  
\bibitem{brownyork}  
J.D.~Brown and J.W.~York,  
\newblock ``Quasilocal energy and conserved  charges derived from the  
gravitational action'',  
\newblock Phys. Rev. D47:1407 (1993),   
  
  
  
\bibitem{juanads}  
J.~Maldacena,  
\newblock ``The large N limit of superconformal field  
theories and supergravity'',  
\newblock Adv. Theor. Math. Phys.2:231-252 (1998), hep-th/9711200.  
  
\bibitem{gkp}  
S.S.~Gubser, I.R.~Klebanov and A.M.~Polyakov,  
\newblock ``Gauge theory correlators from noncritical string theory'',  
\newblock Phys. Lett. B428:105 (1998), hep-th/9802109.  
  
\bibitem{holowit}  
E.~Witten,  
\newblock ``Anti-de Sitter space and holography'',  
\newblock Adv. Theor. Math. Phys.2:253 (1998), hep-th/9802150.   
  
\bibitem{navarro}  
J.~Navarro-Salas and P.~Navarro,  
\newblock ``A note on Einstein gravity on $\ads{3}$ and boundary  
conformal field theory'',  
\newblock Phys. Lett. B439:262 (1998), hep-th/9807019.  
  
\bibitem{emilconf}  
E.J.~Martinec,  
\newblock ``Conformal field theory, geometry and entropy'',  
\newblock hep-th/9809021.  
  
\bibitem{horitzh}  
G.T.~Horowitz and N.~Itzhaki,  
\newblock ``Black holes, shock waves, and causality in the AdS/CFT  
correspondence'',  
\newblock hep-th/9901012.  
  
  
\bibitem{henskend}  
M.~Henningson and K.~Skenderis,  
\newblock ``The holographic Weyl anomaly'',  
\newblock JHEP 9807:023 (1998), hep-th/9806087.  
  
  
\bibitem{hyunetal}  
S.~Hyun, W.T.~Kim and J.~Lee,  
\newblock ``Statistical entropy and AdS/CFT correspondence in BTZ black  
holes'',  
\newblock hep-th/9811005.  
  
\bibitem{chalmers}  
G.~Chalmers and K.~Schalm,  
\newblock ``Holographic normal ordering and multiparticle states in  
the AdS/CFT correspondence'',  
\newblock hep-th/9901144.  
  
\bibitem{odint}  
S.~Nojiri and S.~Odintsov,  
\newblock ``Conformal anomaly for dilaton coupled theories from  
AdS/CFT correspondence'',  
\newblock Phys. Lett. B444:92 (1998), hep-th/9810008.  
  
  
\bibitem{abbdes}  
L.F.~Abbott and S.~Deser,  
\newblock ``Stability of gravity with a cosmological constant'',  
\newblock Nucl. Phys. B195:76 (1982).  
  
\bibitem{ashmag}  
A.~Ashtekar and A.~Magnon,  
\newblock ``Asymptotically anti-de Sitter spacetimes'',  
\newblock Class. Quant. Grav. 1:L39 (1984).  
  
\bibitem{henn}  
M.~Henneaux and C.~Teitelboim,  
\newblock ``Asymptotically anti-de Sitter spaces'',  
\newblock Comm. Math. Phys. 98:391 (1985).  
  
  
\bibitem{mannetal}  
J.D.~Brown, J.~Creighton and R.B.~Mann,  
\newblock ``Temperature, energy and heat capacity of asymptotically  
anti-de Sitter black holes'',  
\newblock Phys. Rev. D50:6394 (1994), hep-th/9405007.  
  
  
  
\bibitem{horhawk}  
G.T.~Horowitz and S.W.~Hawking,  
\newblock ``The gravitational Hamiltonian, action, entropy and surface  
terms'',   
\newblock Class. Quant. Grav.13:1487 (1996), gr-qc/9501014.  
  
\bibitem{hormyers}  
G.T.~Horowitz and R.C.~Myers,  
\newblock ``The AdS/CFT correspondence and a new positive energy  
conjecture for general relativity'',  
\newblock Phys. Rev. D59:026005 (1999), hep-th/9808079.  
  
\bibitem{brownhen}  
J.D.~Brown and M.~Henneaux,  
\newblock ``Central charges in the canonical realization of asymptotic  
symmetries: an example from three-dimensional gravity'',  
\newblock Comm. Math. Phys.104:207 (1986).  
  
\bibitem{weinberg}
S.~Weinberg,
\newblock  {\it Gravitation and Cosmology},
\newblock John Wiley and Sons (1972).  
  
\bibitem{Wald}  
R.M.~Wald,  
\newblock {\it General relativity},  
\newblock University of Chicago Press (1984).  
  
  
  
  
\bibitem{BTZ}  
M.~Ba\~nados, C.~Teitelboim and J.~Zanelli,  
\newblock The black hole in three-dimensional space-time,  
\newblock Phys. Rev. Lett. 69:1849 (1992), hep-th/9204099.  
  
\bibitem{bthz}  
M.~Ba\~nados, M.Henneaux, C.Teitelboim and J.~Zanelli,  
\newblock ``Geometry of the $2+1$ black hole'',  
\newblock Phys. Rev. D48:1506 (1993), gr-qc/9302012.  
  
\bibitem{banados}  
M.~Ba\~nados,  
\newblock ``Global charges in Chern-Simons field theory and the  
$(2+1)$ black hole'',  
\newblock Phys. rev. D52:5816 (1996), hep-th/9405171.  
  
\bibitem{SussWitt}  
L.~Susskind and E.~Witten,  
\newblock ``The holographic bound in anti-de Sitter space'',  
\newblock hep-th/9805114.  
  
  
\bibitem{feff}  
C.~Fefferman and C.R.~Graham, ``Conformal Invariants'', in {\em Elie  
Cartan et les Math\'{e}matiques d'aujourd'hui} (Ast\'{e}risque, 1985) 95.  
  
\bibitem{birrdav}  
N.D.~Birrell and P.C.W.~Davies,  
\newblock {\it Quantum fields in curved space},  
\newblock Cambridge University Press (1982).  
  
\bibitem{gubser}
S.~S.~Gubser, I.~R.~Klebanov and A.~W.~Peet,
``Entropy and Temperature of Black 3-Branes,''
Phys. Rev.  D54, 3915 (1996),
hep-th/9602135.
  
\bibitem{kallraj}  
R.~Kallosh and A.~Rajaraman,  
\newblock ``Vacua of M  theory and string theory'',  
\newblock Phys. Rev. D58:125003, 1998.  
  

\bibitem{probes}
V.~Balasubramanian, P.~Kraus, A.~Lawrence and S.~P.~Trivedi,
 ``Holographic probes of anti-de Sitter space-times,''
Phys. Rev.  D59 , 104021 (1999),
hep-th/9808017.
 
\bibitem{toappear}
V.~Balasubramanian and P.~Kraus,
``Spacetime and the holographic renormalization group,''
Phys. Rev. Lett.   83, 3605 (1999),
hep-th/9903190.  
  
\end{thebibliography}
\end{document}